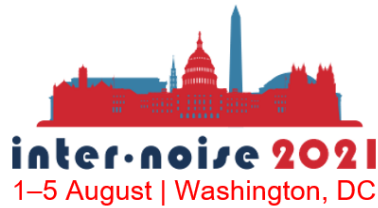

# Effects of aural and visual combinations on appropriateness ratings of residential spaces in an urban city.

Johann Kay Ann, Tan[1], Siu-Kit, Lau[2], Yoshimi Hasegawa[3]
National University of Singapore
Department of Architecture, National University of Singapore, 4 Architecture Drive, Singapore 117566

## ABSTRACT

*This study investigates the aural and visual factors that influence appropriateness perception in soundscape evaluations in residential spaces, where people may spend most of their time in. Appropriateness in soundscape is derived from the expectation of sound sources in a specific environment, place or function heard by a listener. Appropriateness of soundscapes in 30 locations in an urban residential environment is investigated with varying landscape, visual and aural elements through a questionnaire. Participants experienced the soundscape in-situ and were asked to evaluate the appropriateness of soundscape as well as the dominance of specific sound sources such as traffic, human activities and birdsongs in the residential space. The effect of type of traffic on appropriateness is also investigated. A strong relationship is found between appropriateness and affective soundscape qualities such as pleasantness, highlighting the importance of considering appropriateness in soundscape research. In audio-visual combination of specific elements and the partial correlation with appropriateness, specific aural sound sources are found to correlate uniquely to appropriateness while controlling for relevant visual elements, whereas visual elements became redundant in its partial correlation to appropriateness. Residents' perception of appropriateness is found to likely be more dependent on the individual visual elements rather than the overall landscape.*

## 1. INTRODUCTION

Soundscape has been defined as "the acoustic environment as perceived or experienced and/or understood by a person or people, in context" (ISO 12913) [1]. The growing body of soundscape research helps to address worsening noise exposure in urban cities which can have detrimental effects on the wellbeing of residents living in such cities. Soundscape implementation in early stages of urban planning can be most effective in reducing noise exposure to residents [2], while soundscape intervention can help moderate the negative effects of noise that are already present in the environment such as traffic, rail and aircraft noise.

[1] johann.tan@nus.edu.sg

[2] slau@nus.edu.sg

[3] yhasegawa@nus.edu.sg

The Swedish Soundscape-Quality Protocol (SSQP) [3] and ISO 12913 soundscape evaluation involved 8 affective attribute scales: pleasant, chaotic, exciting, eventful, calm, annoying, uneventful, monotonous. [4] Further research has argued for the inclusion of appropriateness and congruency in soundscape literature because of the importance of context in soundscape evaluations. [5][6] The context of the environment may depend on its functions (work, play, live), people, activities, time, landscape and/or place. For example, findings from an experience sampling method study showed that participants are in a better mood in recreational and entertainment soundscapes [7]. When people participate in different activities or spend time in a place (working or studying), there will be different expectation of what the soundscape would be like. The congruency between the expectation and the actual soundscape may influence the pleasantness and value of the soundscape as well as its acoustic comfort, especially in urban soundscape environments. In this study, the appropriateness of soundscapes is studied in the residential context of an urban city, aural and visual sources are investigated as well as the effect of their audio-visual combinations effect on appropriateness ratings.

## 2. METHODOLOGY
### 2.1 Participants

11 participants (8 male and 3 female) took part in this study, the participants visited the residential sites with varying visual landscape and traffic types in Singapore. The participant's age ranged between 21 to 27 years old and self-reported normal hearing and vision and instructed to experience the soundscape for 15 minutes and attempt a questionnaire online.

### 2.2 Sites

The study was conducted in-situ at existing urban residential spaces during daytime (9am – 6pm) while the participants observed the environment for at least 15 minutes and appraised the soundscape based on its appropriateness. 30 locations were identified to represent different landscape (greenery, building, waterbody) and different traffic conditions (heavy, light, none) as well as auditory sources like birdsongs, human activities and mechanical sounds. An online questionnaire was used to collect the subjective responses of appropriateness, overall soundscape quality (OSQ), affective soundscape qualities and the perceived occurrence of sound sources as well as the dominance of visual elements in the environment based on a 5-point Likert scale.

### 2.3 Statistical analysis

The statistical analysis in this paper are performed using the statistical software package, SPSS (v26, IBM, USA). Kolmogorov-Smirnov test is performed to verify the normality of the data collected from participant's questionnaire responses. The data is analyzed using Spearman's rho correlation coefficient, $r_s$ in order to investigate the relationship between the perceived sound sources, visual elements and appropriateness ratings experienced by the participants in the soundscape. A Spearman's rho partial correlation is also performed on the relationship of appropriateness and audio-visual combinations. A two-way repeated-measures analysis of variance (ANOVA) is performed to investigate the main effects of traffic and landscape features on appropriateness. Finally, a Kruskal-Wallis H test was performed on the rank-based non-parametric questionnaire data to discover significant differences in appropriateness ratings.

# 3. RESULTS
## 3.1 Association between sound sources and visual elements.

Appropriateness is derived from the expectations of the listener based on the visual and aural elements perceived, which can influence soundscape quality as experienced, especially in urban soundscape environments. [8] Within urban residential spaces, the degree of visual elements differ for each site usually with a single dominant visual element in the landscape such as elements of waterbody, buildings or greenery. In order to investigate the associations between sound sources and visual elements, Spearman's rho correlation coefficient of each sound source and visual elements are analyzed. The perceived traffic noise is found to correlate strongly and significantly with the visual element 'roads' ($r_s = .905, p < 0.01$) and 'vehicles' ($r_s = .939, p < 0.01$) while negatively correlated with perceived visual greenery ($r_s = -.505, p < 0.01$), 'sky' ($r_s = -.496, p < 0.01$) and 'people' ($r_s = -.428, p < 0.05$). This suggests that the perceived traffic noise have a strong relationship with the visual elements of roads and vehicles, which is within expectations. On the other hand, there is negative correlation strength between the perceived traffic noise and other visual elements such as greenery, sky and people, where the most significant negative correlation is found with greenery and sky. Traffic noise is considered to be annoying to humans and found to cause adverse health effects during long noise exposure [9], reducing the visual element of roads and vehicles through urban and road infrastructure planning can help to moderate the annoyance.

Perceived biophony (birdsongs and insects) correlate negatively with visual element 'roads' ($r_s = -.534, p < 0.01$) and 'vehicles' ($r_s = -.512, p < 0.01$) and positively correlate with 'greenery' ($r_s = .602, p < 0.01$). This shows that biophony is associated with greenery spaces but not with visual traffic factors from vehicles and the road. This is also true for perceived human sounds (activities) such as walking, cycling, conversations, etc, where human sounds are negatively correlated with the visual elements of roads and vehicles ($r_s = -.663, p < 0.01$) and ($r_s = -.666, p < 0.01$) respectively. On the other hand, human sounds correlate positively with the sight of people in the soundscape ($r_s = .783, p < 0.01$), the appropriateness of human sounds is generally influenced by the presence of humans. Finally, the perceived geophonic sounds (wind, trees rustling, water sounds) are significantly correlated with the visual element of water such as lake, pond, or reservoir ($r_s = .497, p < 0.01$) indicating that these geophonic sounds would be appropriate in landscapes with waterbodies. The results also showed a significant negative correlation to visual element of roads and vehicles ($r_s = -.736, p < 0.01$) and ($r_s = -.760, p < 0.01$) respectively. This inverse relationship implies the perception of geophonic sounds may not generally accompany traffic noise in the soundscape of urban residential spaces. Table 1 summarises the results in this section.

Table 1. Spearman's correlation coefficient between sound sources and visual elements

|  | Visual Vehicles | Visual Roads | Visual People | Visual Buildings | Visual Greenery | Visual Sky | Visual Water |
|---|---|---|---|---|---|---|---|
| Perceived Traffic Sounds | .939** $p = .000$ | .905** $p = .000$ | -.428** $p = .018$ | .143 $p = .452$ | -.505** $p = .004$ | -.495** $p = .005$ | -.296 $p = .113$ |
| Perceived Human Sounds | -.666** $p = .000$ | -.663** $p = .000$ | .783** $p = .000$ | .025 $p = .894$ | .313 $p = .092$ | .217 $p = .250$ | .316 $p = .089$ |
| Perceived Biophony Sounds | -.512** $p = .004$ | -.534** $p = .002$ | -.068 $p = .721$ | .197 $p = .297$ | .602** $p = .000$ | -.200 $p = .289$ | .070 $p = .715$ |
| Perceived Geophonic Sounds | -.760** $p = .000$ | -.736** $p = .000$ | .359 $p = .052$ | -.021 $p = .913$ | .416* $p = .022$ | .450* $p = .013$ | .497** $p = .005$ |

\*\* . Correlation is significant at the 0.01 level (2-tailed).* . Correlation is significant at the 0.05 level (2-tailed).
N = 30

### 3.2 Appropriateness of perceived sound sources in residential spaces.

Appropriateness ratings correlated significantly with perceived affective qualities of the soundscape as experienced such as pleasantness ($r_s = .826, p < 0.01$), calm ($r_s = .927, p < 0.01$) and vibrancy ($r_s = .836, p < 0.01$) as well as the overall soundscape quality, OSQ ($r_s = .840, p < 0.01$). This suggests that there is a strong relationship between the way humans perceive the affective quality of the soundscape and the perceived appropriateness of the environment. This highlights the importance of considering the appropriateness of visual and aural factors in the planning and design of soundscape. A soundscape that is appropriate and congruent can help to improve the pleasantness of the environment.

The appropriateness of sound sources in the environment are investigated with Spearman's rho correlation. In terms of sound sources in the 30 residential sites investigated, the perceived occurrence of traffic sounds negatively correlated with appropriateness ($r_s = -.885, p < 0.01$), showing that the presence of traffic sounds in the residential setting is inappropriate as pointed out by previous studies on the exposure of traffic noise to the community. [10] On the other hand, the sounds of birdsongs, human activities and geophonic (wind) correlated positively with appropriateness ($r_s = .547, p < 0.01$), ($r_s = .633, p < 0.01$) and ($r_s = .809, p < 0.01$) respectively.

Table 2. Spearman's correlation coefficient of sound sources to appropriateness

| All landscapes | Perceived Traffic Sounds | Perceived Human Sounds | Perceived Biophony Sounds | Perceived Geophonic Sounds |
|---|---|---|---|---|
| Appropriateness | -.885** $p = .000$ | .633* $p = .000$ | -.547** $p = .002$ | .809** $p = .000$ |

\*\* . Correlation is significant at the 0.01 level (2-tailed).* . Correlation is significant at the 0.05 level (2-tailed).
N = 30

However, the sound sources that influence the perception of appropriateness may also be dependent on the landscape that accompanies the sound sources, as the expectations of sound sources in certain landscapes may influence the appropriateness of different sounds that are heard.

In regards to residential spaces that are dominated by greenery landscape, Spearman's correlation coefficient revealed that the perceived occurrence of traffic noise is indeed negatively correlated with appropriateness ($r_s = -.847$, $p < 0.01$). In residential spaces that have waterbody landscape, the perceived occurrence of traffic noise in such landscape is deemed to be negatively correlated with appropriateness ($r_s = -.746$, $p < 0.05$) but less strongly when compared to greenery spaces. This may be due to higher sound levels of natural sounds around waterbody spaces such as water sounds and birdsongs compared to greenery spaces, especially when the level of non-discrete water sounds are similar to or not less than 3 dB below the urban noise contributed from the traffic. [11] The natural sounds may have moderated the appropriateness ratings as perceived by people when the space is situated near a waterbody as compared to a greenery space which usually lacks natural water sounds. However, it is still possible to use an appropriate water feature installations such as fountains or artificial streams to achieve a similar effect. [12] Separately, in residential spaces that have a greenery landscape, the perceived occurrence of geophonic sounds was correlated with appropriateness ($r_s = .675$, $p < 0.05$). For residential spaces with building features, the perceived occurrence of human activities and biophonic also positively correlated with appropriateness ($r_s = .681$, $p < 0.05$) and ($r_s = .717$, $p < 0.05$) respectively. Interestingly, the perceived occurrence of traffic noise is the most negatively correlated to appropriateness in spaces where buildings are the main visual landscape feature ($r_s = -.933$, $p < 0.01$) indicating that the lack of greenery and waterbody does not necessarily mean that residents are more inclined to tolerate traffic noise. Additionally, unlike the greenery or waterbody spaces, a heavily urbanised estate filled with buildings do not contain as many natural sounds to moderate the traffic noise.

## 4. PARTIAL CORRELATION OF AUDIO-VISUAL COMBINATIONS

It could be argued that between visual elements and aural sound sources, the aural sources take precedence in determining if an audio-visual experience is congruent or appropriate. A theory is that while people can shut their eyes to visual elements or avoid looking at specific objects, they cannot close their ears to unwanted or undesirable sounds, neither can they be selective with the sounds that they hear as the ears pick up on everything that is audible without the ability to 'look away'. Thus, with that in mind, the relationship between audio-visual elements can be considered to be determined by the aural sound sources foremostly. A non-parametric Spearman's partial correlation along with Spearman's zero-order correlation is conducted between relevant sound sources and appropriateness ratings while controlling for relevant visual elements and compared to zero-order correlation results in order to investigate the exclusive relationship between the sound sources and appropriateness. The results show that in almost all cases of audio-visual combination, the Spearman's partial correlations of specific sound sources to appropriateness ratings are significant ($p < 0.05$) when controlled for relevant perceived visual elements. On the other hand, the Spearman's partial correlation of visual elements to the appropriateness are not significant when controlled for relevant perceived sound sources even though the Spearman's zero-order correlation was initially significant for those cases of perceived visual elements to appropriateness ratings. For example, in regards to traffic-related factors, the initial Spearman's zero-order correlation showed that the perceived traffic sounds, the visual element of roads and visual elements of vehicles were all significantly and negatively correlated to the appropriateness ratings ($p < 0.01$) as shown in Table 3. However, while the Spearman's partial correlation between the perceived traffic sounds and appropriateness ratings remain statistically significant after controlling for perceived visual roads and visual vehicles ($r = -.502$, $p < 0.01$), the Spearman's partial correlation between perceived visual element of road and vehicles are no longer statistically significant when controlling for perceived traffic sounds ($r_s = .062$, $p > 0.05$). The conclusion from this analysis is that the original Spearman's zero-order correlation between visual elements of traffic (both road and vehicles individually) and appropriateness can be accounted for in perceived traffic sounds. When the perceived traffic sounds are statistically controlled, the visual elements of traffic no longer correlate significantly with appropriateness. This suggests that the visual elements of traffic became redundant and did not correlate uniquely to appropriateness when the traffic sounds are accounted for. The results are similar for some other audio-visual combination and their partial correlations to appropriateness such as the perceived human sounds with visual people and the perceived geophonic sounds with visual water. The exception to the findings are only present in the geophonic sounds with visual sky and the biophonic sounds with visual greenery, this exception may be due to other unseen factors moderating on appropriateness ratings in these audio-visual combination scenarios. An explanation for the audio-visual combination of geophonic sounds with visual sky may be due to the geophonic sounds that are associated with the sky only constituted the wind, but sound levels of the wind may be too low to be given attention to and thus, the geophonic wind sounds do not uniquely correlate with appropriateness ratings while considering the audio-visual combination with the sky. In regards to biophonic sounds and visual greenery, the relationship between them are not direct like in the case of other audio-visual combination, since biophony sounds originate from the wildlife living in greenery (birds and insects). The results are summarized in Table 4.

Table 3. Spearman zero-order correlation for perceived audio and visual elements

| | Spearman's zero-order correlation | | | | | |
|---|---|---|---|---|---|---|
| | Perceived Visual Vehicles | Perceived Visual Roads | Perceived Visual Greenery | Perceived Visual People | Perceived Visual Sky | Perceived Visual Water |
| Appropriateness | -.841** $p = .000$ | -.798** $p = .000$ | .521** $p = .003$ | .474** $p = .008$ | .580** $p = .001$ | .435** $p = .016$ |
| | Perceived Traffic Sounds | Perceived Biophonic Sounds | Perceived Humans Sounds | Perceived Geophonic Sounds | | |
| Appropriateness | -.885** $p = .000$ | .547** $p = .002$ | .633** $p = .000$ | .809** $p = .000$ | | |

** . Correlation is significant at the 0.01 level (2-tailed).* . Correlation is significant at the 0.05 level (2-tailed).
N = 30

Table 4. Spearman partial correlation for audio-visual combination

| Spearman's partial correlation | | | | |
|---|---|---|---|---|
| | **Perceived Traffic Sounds** | | **Visual Road** | **Visual Vehicle** |
| Control variable | | Control variable | | |
| Visual Road & Visual Vehicles | -.502** p = .007 | Perceived Traffic Sounds | .019 p = .920 | -.062 p = .750 |
| | **Perceived Human Sounds** | | **Visual People** | |
| Control variable | | Control variable | | |
| Visual People | .477** p = .009 | Perceived Human Sounds | -.043 p = .823 | |
| | **Perceived Geophonic Sounds** | | **Visual Water** | **Visual Sky** |
| Control variable | | Control variable | | |
| Visual Water & Visual Sky | .747** p = .000 | Perceived Geophonic Sounds | .065 p = .739 | .411** p = .027 |
| | **Perceived Biophonic Sounds** | | **Visual Greenery** | |
| Control variable | | Control variable | | |
| Visual Greenery | .343 p = .069 | Perceived Biophonic Sounds | .287 p = .131 | |

\*\* . Correlation is significant at the 0.01 level (2-tailed). \* . Correlation is significant at the 0.05 level (2-tailed).
N = 30

## 5. THE EFFECT OF VARIED TRAFFIC CONDITIONS AND LANDSCAPE FEATURES ON APPROPRIATENESS

In the previous sections, Spearman's correlation identified the correlation strength of traffic sound sources and traffic-related visual elements to have the strongest correlations strength to appropriateness. The traffic type of heavy generally had the lowest appropriateness ratings despite of different landscape features while light traffic only slightly affected appropriateness ratings as compared to no traffic and generally did not drop more than 1-point as shown in Figure 1 where the mean appropriateness ratings of each site are represented in the chart grouped by traffic conditions.

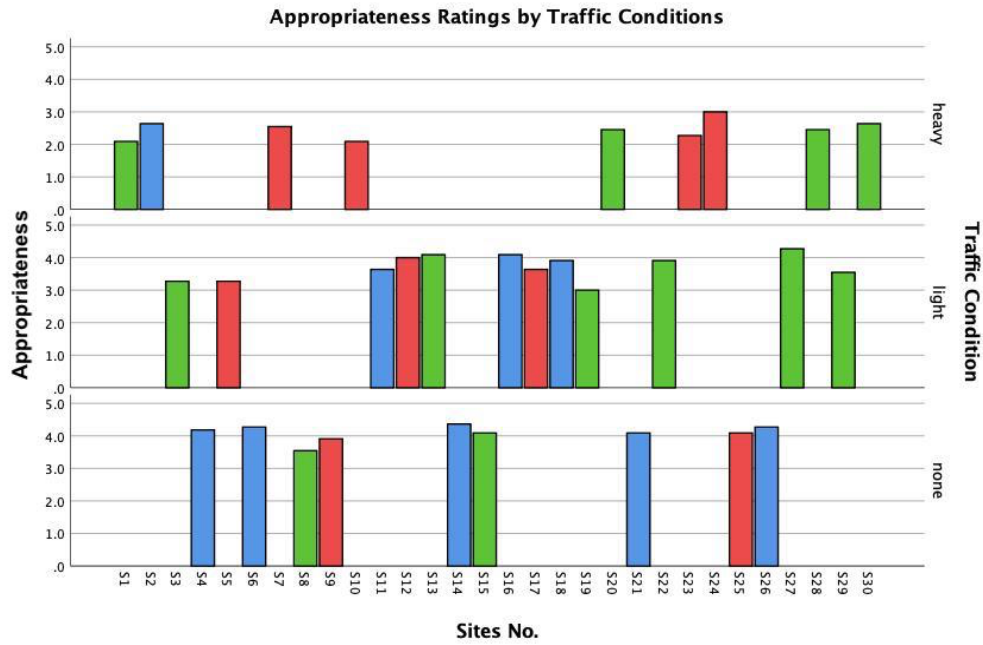

Figure 1. Appropriateness ratings grouped by traffic conditions

A two-way repeated-measures analysis of variance (ANOVA) is performed on appropriateness ratings taking into consideration different combinations of traffic types and landscape features. The result shows that only traffic types have a main effect on appropriateness [$F(2,20) = 37.564$, $p < 0.01$, $\eta_p^2 = 0.790$] (Table 5) while landscape features did not have a significant main effect on appropriateness [$F(1.296, 12.963) = 1.620$, $p = 0.231$, $\eta_p^2 = 0.139$]. There is also no significant interaction between traffic conditions and landscape features on appropriateness ratings [$F(4, 40) = 0.493$, $p = 0.741$, $\eta_p^2 = 0.047$]. Figure 2 shows the estimated marginal means of appropriateness ratings in different traffic conditions which indicates a downward trend in appropriateness from none (no traffic) to light traffic and heavy traffic. In spaces that are exposed to traffic, landscape features did not influence the appropriateness ratings except for in the case of waterbody which was only slightly higher in appropriateness ratings. This suggests that the participant's perception of appropriateness may be more dependent on individual visual elements in the space rather than the overall landscape, this hypothesis is investigated in Section 6.

Table 5. Summary of ANOVA results for Appropriateness ratings

| Factors of Appropriateness | $df_1$ | $df_2$ | F | p | $\eta_p^2$ |
|---|---|---|---|---|---|
| Traffic Conditions | 2 | 20 | 37.564 | < 0.001 | 0.790 |
| Landscape Features [a] | 1.296 | 12.963 | 1.620 | 0.232 | 0.139 |
| Traffic * Landscape | 4 | 40 | 0.493 | 0.741 | 0.047 |

[a] Assumption of sphericity was violated and Huynh-Feldt correction was applied

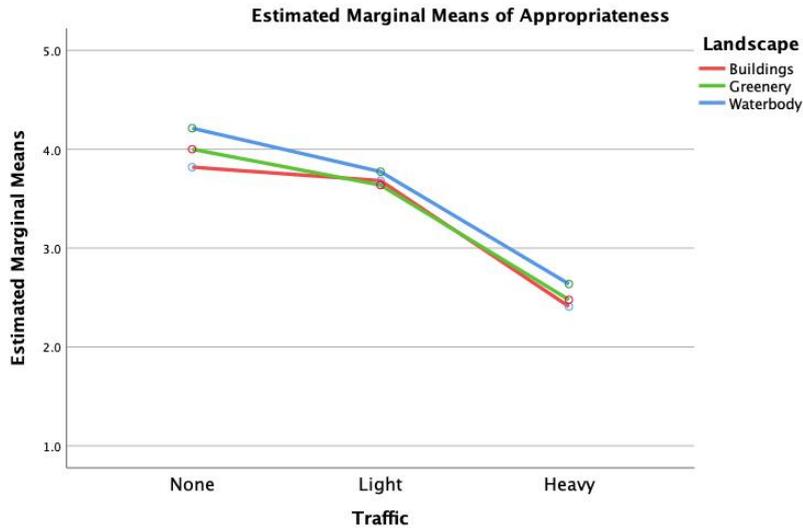

Figure 2. Estimated marginal means of Appropriateness

The estimated marginal means of appropriateness are also shown according to traffic conditions and landscape features. Between different traffic conditions are reflected in Table 6 while the comparison of means between landscape features are reflected in Table 7. The means for traffic conditions showed a larger difference between the mean scores where the lowest appropriate scores are when the traffic condition is heavy (Appropriateness = 2.508 ± 0.163) whereas the estimated marginal means for landscape features have less variance. (Table 7)

Table 6. Estimated Marginal Means (Traffic conditions)

| Traffic Condition | Mean Appropriateness | Std. Error | 95% Confidence Interval | |
|---|---|---|---|---|
| | | | Lower Bound | Upper Bound |
| None | 4.010 | 0.085 | 3.820 | 4.200 |
| Light | 3.697 | 0.106 | 3.461 | 3.933 |
| Heavy | 2.508 | 0.163 | 2.144 | 2.871 |

Table 7. Estimated Marginal Means (Landscape feature)

| Landscape features | Mean Appropriateness | Std. Error | 95% Confidence Interval | |
|---|---|---|---|---|
| | | | Lower Bound | Upper Bound |
| Greenery | 3.303 | 0.078 | 3.129 | 3.477 |
| Building | 3.371 | 0.063 | 3.232 | 3.511 |
| Waterbody | 3.540 | 0.142 | 3.225 | 3.856 |

## 6. THE EFFECT OF INDIVIDUAL VISUAL ELEMENT IN THE LANDSCAPE ON APPROPRIATNESS

In regards to the perceived individual visual element in the landscape, the participant's perception of appropriateness may be accounted for by individual visual elements rather than the overall landscape as the estimated marginal means of different landscape features had little variance to appropriateness. (Table 7) Thus, a Kruskal-Wallis H test is conducted with the median scores of appropriateness and median scores of perceived visual elements to examine if there are a statistically significant difference in different levels of perceived appropriateness. The results show that in the case of the perceived visual elements of roads ($\chi^2(3) = 15.225$, $p = 0.002$), greenery ($\chi^2(3) = 10.038$, $p = 0.018$), and people ($\chi^2(2) = 6.931$, $p = 0.031$), there are statistically significant differences in appropriateness ratings. However, there are no statistically significant differences in appropriateness ratings for the perceived visual elements of the sky, water and buildings. It could be due to having perceived visual sky being dominant in all the cases in this study. The study also generally involved some buildings in every site because the study focused on residential spaces. Thus, the study concludes that the visual elements of roads, greenery and people in the overall landscape account for the perception of appropriateness ratings.

The Kruskal-Wallis H and mean rank appropriateness ratings for the different individual perceived visual element are summarized in Table 8.

Table 8. Kruskal-Wallis H test for individual perceived visual elements in appropriateness ratings

| Perceived Visual Roads Score | N | Mean rank | Appropriateness | | Asymp. Sig. |
|---|---|---|---|---|---|
| | | | Kruskal-Wallis H | df | |
| 1 | 7 | 21.50 | | | |
| 2 | 10 | 18.40 | 15.225 | 3 | .002 |
| 3 | 7 | 13.50 | | | |
| 4 | 6 | 6.00 | | | |
| Total | 30 | | | | |

| Perceived Visual Greenery Score | N | Mean rank | Appropriateness | | Asymp. Sig. |
|---|---|---|---|---|---|
| | | | Kruskal-Wallis H | df | |
| 2 | 3 | 7.00 | | | |
| 3 | 5 | 9.10 | 10.038 | 3 | 0.018 |
| 4 | 20 | 17.77 | | | |
| 5 | 2 | 21.50 | | | |
| Total | 30 | | | | |

| Perceived Visual People Score | N | Mean rank | Appropriateness | | Asymp. Sig. |
|---|---|---|---|---|---|
| | | | Kruskal-Wallis H | df | |
| 2 | 19 | 12.68 | | | |
| 3 | 10 | 20.25 | 6.931 | 2 | 0.031 |
| 4 | 1 | 21.50 | | | |
| Total | 30 | | | | |

## 7. CONCLUSION

The strong relationship between perceived appropriateness and affective quality of the soundscape affirms the importance of appropriateness and congruency in soundscape planning, design, implementation and intervention in the context of living spaces in residential areas. An appropriate soundscape can help to improve the pleasantness of the environment. For example, in living spaces, traffic noise correlated negatively with appropriateness rating which was true for all landscape feature even for urbanized spaces where buildings were the main landscape feature, indicating that residents are not more inclined to tolerate traffic noise in heavily urbanized spaces without other natural features. This highlights the importance of having some natural features such as greenery, the introduction of birdsongs and artificial water features to compensate for the lack of natural sounds in a heavily urbanized living spaces.

In regards to aural versus visual elements, Spearman's partial correlation revealed that generally for most cases of audio-visual combination in the present study, the partial correlation of specific sound sources to appropriateness ratings are still statistically significant while controlling for relevant visual elements. Visual elements on the other hand were not found to correlate uniquely to appropriateness while controlling for the relevant sound sources which indicated in the study that visual elements became redundant and did not correlate uniquely to appropriateness while the relevant sound sources were accounted for.

Between the effect of traffic conditions and landscape features on appropriateness ratings, two-way repeated-measures ANOVA identified a statistically significant main effect of traffic conditions on appropriateness. The landscape features did not have any significant main effect and the estimated marginal means for varying landscape features (greenery, buildings, waterbody) did not show a sizeable variance in appropriateness ratings. This led the study to compare the rank score of perceived dominance of visual elements individually in the landscape instead of the overall landscape and how it affects appropriateness ratings. The use of the Kruskal-Wallis H test indicated that there were indeed statistically significant differences in appropriateness ratings for perceived dominance of individual visual elements. The study concludes that residents' perception of appropriateness is likely dependent on the individual visual element in the space instead of considering the overall landscape. Thus, this supports the previous recommendation for urban planners to consider the addition of natural features in heavily urbanized spaces to avoid inappropriateness of the soundscape in the context of living spaces that has been exacerbated by artificial elements such as traffic and other undesired sounds.

## 8. ACKNOWLEDGEMENTS

We gratefully acknowledge the Ministry of Education (MOE), Singapore, for supporting this research project under its Academic Research Fund Tier 2 (MOE2018-T2-1-105).